\documentclass[12pt]{article}

\usepackage[margin=1in]{geometry}
\usepackage[version=3]{mhchem}
\usepackage{graphicx}
\usepackage{url}
\usepackage{float}
\usepackage{amsfonts,amssymb,amsmath}
\usepackage{booktabs}
\usepackage{multirow}
\usepackage{subcaption}
\usepackage[font=footnotesize,labelfont=bf]{caption}
\usepackage{placeins}
\usepackage{xcolor}
\usepackage{scalerel}
\usepackage{makecell}
\usepackage{custom_commands}
\usepackage[super,sort&compress]{natbib}
\usepackage{hyperref}
\usepackage[capitalize]{cleveref}
\usepackage[affil-it]{authblk}

\makeatletter
\renewcommand\paragraph{\@startsection{paragraph}{4}{\parindent}%
  {1ex \@plus 0.25ex \@minus 0.12ex}%
  {-1em}%
  {\normalfont\normalsize\itshape}}
\makeatother
\graphicspath{{figures/}}

\title{Enhancing molecular dynamics with equivariant machine-learned densities}

\author[1]{Mihail Bogojeski}
\author[2]{Muhammad R. Hasyim}
\author[2]{Leslie Vogt-Maranto}
\author[1,3,4]{Klaus-Robert M\"uller\thanks{klaus-robert.mueller@tu-berlin.de}}
\author[5,6]{Kieron Burke\thanks{kieron@uci.edu}}
\author[2,7,8]{Mark E. Tuckerman\thanks{mark.tuckerman@nyu.edu}}

\affil[1]{BIFOLD and Machine Learning Group, Technische Universit\"at Berlin, Franklinstr.\ 28/29, 10587 Berlin, Germany}
\affil[2]{Department of Chemistry, New York University, New York, NY 10003, USA}
\affil[3]{Department of Artificial Intelligence, Korea University, Seoul 02841, Korea}
\affil[4]{Max-Planck-Institut f\"ur Informatik, 66123 Saarbr\"ucken, Germany}
\affil[5]{Department of Physics and Astronomy, University of California, Irvine, CA 92697, USA}
\affil[6]{Department of Chemistry, University of California, Irvine, CA 92697, USA}
\affil[7]{Courant Institute of Mathematical Sciences, New York University, New York, NY 10012, USA}
\affil[8]{NYU-ECNU Center for Computational Chemistry at NYU Shanghai, 3663 Zhongshan Road North, Shanghai 200062, China}

\date{}

\begin{document}

\maketitle

\begin{abstract}
Machine-learning interatomic potentials (MLIPs) have enabled molecular dynamics at near \textit{ab initio} accuracy, yet remain limited to energies and forces by construction, leaving electronic observables such as dipole moments and polarizabilities inaccessible.
We introduce DenSNet, a density-first approach to machine-learned electronic structure that learns the Hohenberg--Kohn map from nuclear configurations to the ground-state electron density.
Our approach employs an SE(3)-equivariant neural network to predict density coefficients of a flexible atom-centered Gaussian basis, combined with a $\Delta$-learning strategy that uses superposed atomic densities as a prior to accelerate training.
A second equivariant network then maps the predicted density to the total energy, providing a unified framework for molecular dynamics and electronic structure.
We validate DenSNet on ethanol, ethanethiol, and resorcinol, where infrared spectra from machine-learned trajectories show excellent agreement with experimental gas-phase measurements.
To test scalability, we train on polythiophene oligomers with 1--6 monomers and extrapolate to chains of up to 12 monomers, generating stable long-time trajectories whose infrared spectra agree with reference density functional theory calculations.
Here, we show that reinstating the electron density as the central learned quantity opens a practical route to transferable prediction of spectroscopic and electronic observables in large-scale molecular simulations.
\end{abstract}

\section{Introduction}\label{sec:intro}
The combination of electronic structure calculations and molecular dynamics (MD) simulations, known as {\it ab initio} MD (AIMD), has had a profound impact on atomistic modeling of chemical dynamics in condensed phases.~\cite{car1985unified,tuckerman2002abinitio,marx2009abinitio}
However, the computational overhead of electronic structure calculations limits routine AIMD to system sizes of order $10^3$ atoms and trajectories of at most a few hundred picoseconds.
To mitigate this cost, machine learning interatomic potentials (MLIPs)~\cite{behler2011atom,schutt2017schnet,unke2021chemrev} have emerged as a practical alternative. MLIPs bypass repeated electronic structure calculations by learning potential energy surfaces from reference data and can predict energies and forces with fidelity approaching that of the underlying {\it ab initio} theory.~\cite{behler2007generalized, chmiela2017machine, batzner2021nequip, batatia2022mace}  
As a result, MLIPs enable simulations at a fraction of the cost of electronic structure calculations while retaining much of the accuracy of first-principles methods.

Despite these advances, MLIPs generally lack direct access to the electronic observables required for spectroscopy, such as the dipole moments and polarizability tensors that determine infrared (IR) absorption and Raman scattering, respectively.~\cite{jindal2025computing,kapil2023firstprinciples} 
Because these quantities depend on the electronic structure, MLIPs are often supplemented with additional models trained to predict selected properties alongside energies and forces.~\cite{ji2025universal} 
Learning each observable separately, however, is neither scalable nor does it provide a unified route to the broad range of properties one may wish to compute.

A more general strategy is motivated by density functional theory (DFT).
The first Hohenberg--Kohn theorem establishes (up to an additive constant) a one-to-one correspondence between the external potential $v(\mathbf r)$, defined by the Coulomb interaction of the electrons with the nuclei, and the nondegenerate ground-state electron density $\rho(\mathbf r)$.~\cite{hohenberg1964inhomogeneous} 
This correspondence motivates learning the map from the nuclear configuration, specified by its charges and coordinates (equivalent to $v({\bf r})$), to the electron density, i.e., $v({\bf r}) \longleftrightarrow \rho({\bf r})$. 
Once an accurate density is available, ground-state properties including total energies, electrostatic potentials, and dipole-related response properties follow directly. 
Early work in this direction applied kernel-based methods to predict molecular atomization energies from nuclear charges and positions~\cite{rupp2012fast} and used machine learning to approximate density functionals for one-dimensional model systems.~\cite{snyder2012finding,li2014understanding}
The ground-state density itself was later learned from nuclear positions via kernel ridge regression, an approach we refer to as the ML--Hohenberg--Kohn (ML-HK) map, which also produced the first molecular dynamics simulation with a machine-learned density.~\cite{brockherde2017bypassing}
Subsequent work applied the ML-HK framework to reach coupled-cluster accuracy through density-based $\Delta$-learning~\cite{bogojeski2020quantum} and to predict excited-state densities~\cite{bai2022machine} and one-particle density matrices, the latter enabling energy-conserving molecular dynamics and infrared spectra from learned electronic structure.~\cite{shao2023machine}
Separately, neural-network density functionals have been trained to address the fractional-electron problem, improving the functional itself rather than bypassing the Kohn--Sham equations.~\cite{kirkpatrick2021pushing}
More recently, atom-centered learning of the electron density has been used to recover energies via a single Kohn--Sham diagonalization step;~\cite{grisafi2022electronic} for a recent review on applying ML for energy functionals, see ref.~\cite{huang2023dft}.

\begin{figure}[!tb]
  \centering\includegraphics[width=\linewidth]{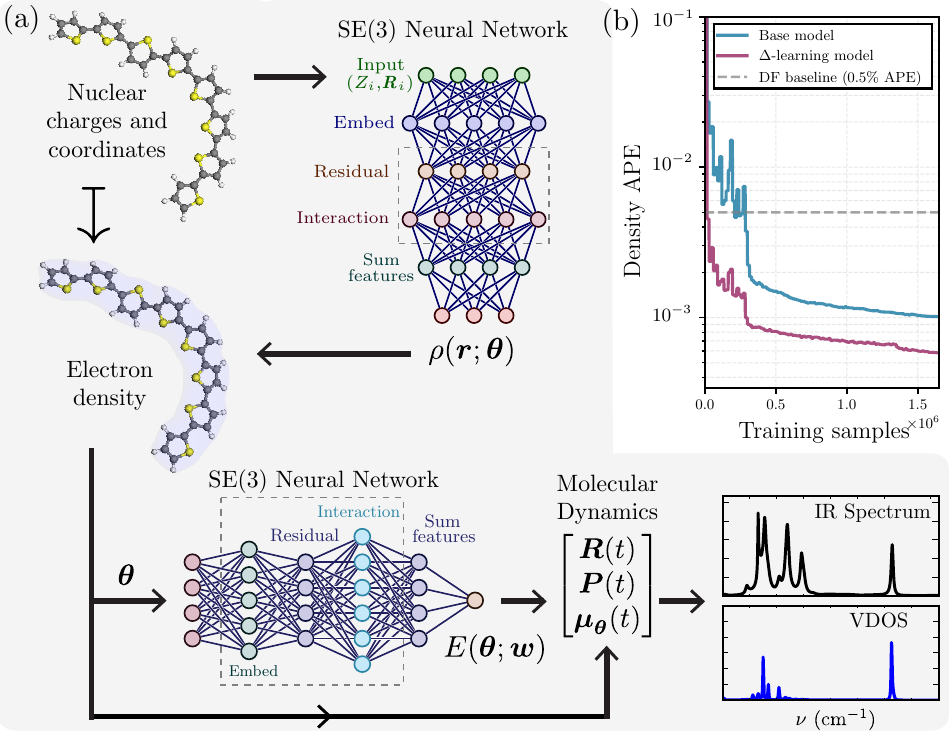}
  \caption{Overview of DenSNet. 
  \textbf{(a)} Two-stage architecture for density-enhanced molecular dynamics, illustrated with a polythiophene oligomer. 
  The first SE(3)-equivariant neural network (top) takes nuclear charges and coordinates as input and predicts electron density coefficients $\rho(\rv;\boldsymbol{\theta})$ in a flexible atom-centered basis. 
  These coefficients then serve as input to a second SE(3)-equivariant network (bottom) that predicts the total energy $E(\boldsymbol{\theta};\mathbf{w})$. 
  Molecular dynamics trajectories provide atomic positions $\mathbf{R}(t)$, momenta $\mathbf{P}(t)$, and dipole moments $\boldsymbol{\mu}_{\boldsymbol{\theta}}(t)$ computed directly from the predicted density, enabling calculation of infrared (IR) absorption spectra and vibrational density of states (VDOS). 
  \textbf{(b)} Training convergence comparison showing density absolute fractional error (AFE) versus training epoch. $\Delta$-learning with superposed atomic densities (SAD) as a prior (magenta) achieves faster convergence and lower final error than learning the full density from scratch (cyan). The dashed line indicates the baseline error of density fitting.}
  \label{fig:overview}
\end{figure}

Existing approaches to learning the ML-HK map, however, face practical challenges. 
Methods that represent the density on a fixed global grid or in a global basis tied to a laboratory-frame coordinate system, e.g., kernel ridge regression, do not automatically respect the transformation properties of $\rho(\mathbf r)$ under rotations and translations of the nuclear configuration. 
As a consequence, one must either realign configurations to the training frame or rely on extensive data augmentation, both of which increase training cost and complicate deployment in long MD simulation runs. 
These fixed-grid representations also hinder scalability to larger systems that share the same chemical motifs because they do not naturally encode locality and extensivity.

In this paper, we address these limitations by constructing a message-passing neural network (MPNN) that is equivariant under the special Euclidean group SE(3), the group of rotations and translations that preserve Euclidean distances.~\cite{thomas2018tensor, weiler20183d}  
In particular, the MPNN models an atom-centered basis expansion of the electron density using products of radial functions and spherical harmonics, with coefficients that transform equivariantly under rotations while the reconstructed density remains a scalar field. 
Our model not only predicts density coefficients but also adapts the radial basis functions in a configuration-dependent manner, yielding a flexible representation that responds to changes in bonding patterns and local electronic environments. 
Coupling the learned density to a companion energy network provides forces for molecular dynamics while retaining access to electronic observables derived from the predicted density; see Figure~\ref{fig:overview}(a). 
We term this density-based equivariant framework \textbf{DenSNet} and demonstrate its capabilities in two ways: (1) computing infrared spectra of ethanol, ethanethiol, and resorcinol from machine-learned trajectories, obtaining excellent agreement with experimental gas-phase measurements, and (2) training on polythiophene oligomers with 1--6 monomers that extrapolate to chains twice as large, yielding stable dynamics and accurate spectra for systems well beyond the training distribution.

\FloatBarrier
\section{Results}\label{sec:results}

In this section, we first introduce the equivariant density model and its atom-centered basis, following which, we benchmark them on a number of small-molecule examples.

\subsection{Machine learning of equivariant electron densities}\label{sec:equiv_dens}

Our SE(3)-equivariant neural network architecture for electron densities builds on PhiSNet.~\citep{unke2021se}
PhiSNet uses spherical harmonic representations that reflect the orientation of the molecule and ensures that all transformations in the network are rotationally equivariant, producing an atom-centered basis representation of the density in the correct orientation.
While the original PhiSNet was designed to predict electronic Hamiltonians and wavefunctions, we designed its core equivariant message-passing modules to directly predict electron density coefficients, introducing several key modifications: (1) a shallower network architecture (three modules instead of five), (2) direct prediction of density basis coefficients rather than matrix elements, (3) $\Delta$-learning with superposed atomic densities (SAD; see below) as a prior, and (4) physical constraints tailored to electron densities (charge conservation and positivity). 
We begin by describing the atom-centered basis used to construct the density, then turn to the neural network that takes the basis coefficients as input features. 

\paragraph{Density basis definition.}\label{sec:ml_dens_basis}
The task of our ML model is to predict the electron density $\rho(\rv)$ of a molecule with $M$ atoms given only the positions $\Rr = \{\Rr_i | i \in 1\dots M\}$ and charges $\mathrm{Z} = \{Z_i | i \in 1\dots M\}$ of the atoms, which we achieved via an atom-centered basis set with learnable weights. 
The density-fitting parameters
$\bomega \equiv \{\alpha,\beta,\gamma\}$ used by our ML density model consist of the radial width parameters $\balpha$, the radial scale coefficients $\bbeta$, and the angular coefficients $\bgamma$, corresponding to the width, scale, and angular (spherical harmonics) components of a Gaussian-type orbital (GTO) density-fitting basis, respectively. The number of angular shells of degree $l$ for atom $i$ is $J_{il}$, a quantity that depends only on the element type~$Z_i$ so all atoms of the same element share the same set of angular channels, e.g., all carbon atoms in a molecule have identical $J_{il}$ for each~$l$. The density is decomposed into atom-centered contributions:
\begin{equation}\label{eq:ml_basis_def}
  \rho_{\bomega}(\rv) = \sum_{i=1}^{M}\rho_i(\rv_i),\qquad \rv_i = \rv - \Rr_i,
\end{equation}
where each atom's contribution is an expansion in real spherical harmonics and Gaussian radial functions:
\begin{equation}\label{eq:atom_density}
  \rho_i(\rv_i) = \sum_{l=0}^{L}\sum_{j=1}^{J_{il}} \bgamma_{ij}^{(l)}\,Y_l(\hat{\rv}_i)\;\beta_{ilj}\,A_{ilj}\,e^{-\alpha_{ilj}\,|\rv_i|^2},
\end{equation}
with $|\rv_i| = \lVert\rv_i\rVert$ and $\hat{\rv}_i = \rv_i/|\rv_i|$. Here $Y_l(\hat{\rv}_i)$ denotes the vector of all $(2l+1)$ real spherical harmonics of degree $l$, and the product $\bgamma^{(l)}_{ij}\,Y_{l}(\hat{\rv}_i)$ is a dot product over the $m$ indices. Each angular shell~$j$ has a single associated radial function (an uncontracted basis), matching the composition of aug-cc-pVQZ-jkfit. The implementation supports contracted bases, but here we use an uncontracted basis by choice. The normalization constant $A_{ilj} = (2\alpha_{ilj}/\pi)^{3/4}\sqrt{(4\alpha_{ilj})^l/(2l-1)!!}$ ensures unit integral of each primitive, where $(2l-1)!! = 1 \cdot 3 \cdot 5 \cdots (2l-1)$ is the double factorial. A step-by-step worked example with explicit index enumeration for a carbon atom is given in Supplementary Note~1.

While the basis presented here resembles a standard GTO density-fitting basis, we note that standard basis sets fix the parameters $\bbeta$ and $\balpha$ to a constant and let the coefficients for the spherical components $\bgamma$ be optimizable. In standard density fitting, these parameters are determined by minimizing the integrated absolute difference between the basis-set expansion $\rho_{\bomega}(\rv)$ and the target density $\rho(\rv)$, yielding the optimization objective
\begin{equation}\label{eq:int_dens_diff}
  \min\limits_{\bomega} \int\left|\rho_{\bomega}(\rv) - \rho(\rv)\right|d\rv.
\end{equation}
In contrast, our basis allows the parameters $\bbeta$ and $\balpha$ to take distinct values for each atomistic system, giving the basis greater expressivity. 

\paragraph{Equivariant density prediction network.}\label{sec:den_pred_net}
The density-fitting parameters
$\bomega \equiv \{\alpha,\beta,\gamma\}$ of the atom-centered basis in Equation~(\ref{eq:ml_basis_def}) are predicted by a neural network built on the PhiSNet architecture~\citep{unke2021se} 
(see Fig.~\ref{fig:nn_architecture}).
Each atom is first assigned an initial feature vector based on its element, encoding nuclear charge and ground-state electronic configuration.
The network then refines these features through three rounds of message passing.

In each round, every atom collects information from its neighbors within a cutoff radius, weighted by the direction and distance of the interatomic separation expressed through spherical harmonics. The gathered messages are combined with the atom's own features via learned nonlinear transformations with skip connections, thereby updating each atom's representation while preserving SE(3) equivariance (see Supplementary Note~1 for detailed definitions of all operations).
After three such rounds, the network produces atom-wise feature vectors
\begin{equation}\label{eq:beta_coeffs}
  \fv_i = \bigl\{\fv_i^{(0)}, \fv_i^{(1)}, \ldots, \fv_i^{(L_{\max})}\bigr\}, \quad i \in \{1,\ldots,M\},
\end{equation}
which encode information about each atom's environment in a spherical tensor structure.

Using these per-atom features, we extract the three basis parameter sets $\{\alpha,\beta,\gamma\}$ as follows.
The radial scale coefficients $\bbeta$ and Gaussian exponents $\balpha$ are
obtained from the rotationally invariant features $\fv_i^{(0)}$ via learned
projections, while the angular coefficients $\bgamma^{(l)}$, which encode the orientational
character of the density, are extracted from spherical harmonic features of the corresponding degree with a special linear layer that mixes angular channels while preserving equivariance. A detailed explanation of this basis parameter extraction can be found in Supplementary Note~1.
With all three sets of parameters in hand, the density can be evaluated in coordinate space via Equation~(\ref{eq:ml_basis_def}). 
For improved convergence, $\balpha$ and $\bbeta$ can be optionally initialized to the values of an existing density fitting basis (see Supplementary Note~2).

This density model, which we call DenSNet, is then trained using stochastic gradient descent by minimizing the absolute fractional error (AFE), i.e., the integrated density error normalized by the total electron count, to the reference density, which is a normalized version of the objective in Equation~(\ref{eq:int_dens_diff}) to ensure that atomistic systems of different sizes are treated equally:
\begin{equation}\label{eq:perc_dens_diff}
  \mathcal{L}_{\mathrm{dens}}(\{\rho_{\bomega,b}\}_{b=1}^{N_{\mathrm{batch}}},\{\rho_b\}_{b=1}^{N_{\mathrm{batch}}}) = \frac{1}{N_{\mathrm{batch}}}\sum_{b=1}^{N_{\mathrm{batch}}}\scaleobj{1.5}{\int} \frac{\left|\rho_{\bomega,b}(\rv) - \rho_b(\rv)\right|}{\int\rho_b(\rv)d\rv}d\rv.
\end{equation}
The integral is numerically evaluated on angular-radial integration grids, which provide accurate estimates of the loss with a smaller, linearly scaling number of grid points (see Supplementary Note~2 for a grid resolution analysis). 

\paragraph{$\Delta$-learning of electron densities.}\label{sec:delta-learning}
The density around the nuclei contains most of the total charge and shows minimal variation across geometries, mainly because it describes the distribution of core electrons, which do not participate in bonding. In contrast, the valence-electron density responsible for chemical bonding varies substantially with molecular geometry, yet it accounts for only a small fraction of the total electron count. This separation of scales suggests that learning the full density from scratch is inefficient, as the model must allocate capacity to represent the largely invariant core region.

We aimed to exploit this property of the density to improve the model's accuracy and transferability. We partitioned the density as follows:
\begin{equation}\label{eq:delta_rho}
\rho(\rv) = \underbrace{\sum_{i=0}^M \rho_{Z_i, \mathrm{free}}}_{\rho_{\mathrm{SAD}}} + \rho_{\Delta\mathrm{ML}},
\end{equation}
where $\rho_{Z_i, \mathrm{free}}$ denotes the atomic density of the atom with charge $Z_i$ and $\rho_{\Delta\mathrm{ML}}$ denotes the machine-learned part of the density as represented in Equation~(\ref{eq:ml_basis_def}). This allows us to use the superpositions of atomic densities (SAD), denoted as $\rho_{\mathrm{SAD}}$, as a baseline for the unchanging part of the density, and use our ML model to learn the changes to the SAD due to bonding. This $\Delta$-learning strategy extends the original PhiSNet framework,\citep{unke2021se} which directly predicts Hamiltonian matrices without a baseline reference. Our proposed approach makes the learning problem simpler by focusing the expressive power of the neural network and the basis representation solely on the smoother residual bonding density, leading to faster convergence and improved accuracy (see Figure~\ref{fig:overview}(b)).~\cite{huang2025active}

The $\Delta$-learning approach also mitigates a separate problem arising from the density-fitting basis. Density-fitting basis sets are often uncontracted and struggle to represent the core region accurately compared to the density obtained from the product of atomic orbitals. By using the actual atomic densities obtained from DFT calculations using the same basis set, we obtained an accurate representation of the core region, thereby eliminating this particular source of error (see Supplementary Note~2 for pre-training procedure details).

\paragraph{Learning densities with constraints.}
Incorporating geometric symmetries and physical constraints is known to improve the accuracy and data efficiency, e.g., GDML/sGDML.~\cite{chmiela2017machine,chmiela2018towards}  
Up to this point, our model incorporated only the geometric symmetries through its equivariance properties. Therefore we additionally modeled the \emph{physical} conditions in our architecture. Specifically, the nature of density as an electron probability distribution also imposes two \emph{physical} conditions. First, the integral of the electron density must be equal to the number of electrons in the system ($\int \rho(\rv) d\rv = N_e$), and second, the density must be nonnegative everywhere ($\rho(\rv) \geq 0$). 
Since the integral of any spherical harmonic function with degree $l > 0$ is zero, we can easily ensure that the integral of our predicted densities is equal to the number of electrons by only constraining the coefficients of the spherical basis components with degree $l=0$. Since the spherical harmonic of degree $0$ is a constant, it can be left out, leaving us only with the Gaussian radial functions $e^{-\alpha_{i0j}\,|\rv_i|^2}$, which can be normalized analytically to set the integral of all atom-centered basis functions with degree $l=0$ to 1. Then, to guarantee that the density integrates to the correct value, we constrain the $\bgamma$ and $\bbeta$ coefficients such that:
\begin{equation}\label{eq:dens_constraint}
  \sum_{i=1}^M \sum_{j=1}^{J_{i0}}\bgamma_{i0j}\,\beta_{i0j} = N_\mathrm{el}.
\end{equation}
While this can be achieved in many ways, we adopted a straightforward approach, namely dividing all relevant coefficients by their sum and multiplying by the number of electrons. This ensures that the integral of the density is always correct. 

When using the $\Delta$-learning model, $\rho_{\mathrm{SAD}}$ already yields the correct integral, allowing us to guarantee the conservation of this constraint by ensuring that the integral of $\rho_{\Delta\mathrm{ML}}$ is equal to zero. For simplicity and without loss of generality, we can assume that all of the $\mathbf{\gamma}$ coefficients for $l=0$ are 1, and let the vector $\bbeta_0$ contain all the scale coefficients $\beta_{i0j}$ corresponding to basis functions with degree $l=0$. Similarly to the density gradient projection in Zhang \textit{et al.},~\cite{zhang2024overcoming} we can transform the $\bbeta$ coefficients of the ML density to project it to the manifold of densities with an integral of 0 as follows:
\begin{equation}\label{eq:integral_proj}
\bbeta_{\mathrm{proj}}=\left(\mathbf{I}-\frac{\mathbf{a} \mathbf{a}^{\top}}{\mathbf{a}^{\top} \mathbf{a}}\right) \bbeta,
\end{equation}
where the vector $\mathbf{a}$ consists of the integrals of all of the radial basis functions corresponding to the $\bbeta$ coefficients:
\begin{equation}\label{eq:norm_vector}
\mathbf{a}_{ij} = A_{i0j} \int_0^\infty 4\pi\, r^2\, Y_0\, e^{-\alpha_{i0j}\, r^2}\, dr = A_{i0j}\,\frac{\pi}{2\,\alpha_{i0j}^{3/2}}, \quad r=\lVert\rv_i\rVert.
\end{equation}
This projection ensures that the combination of $\rho_{\mathrm{SAD}}$ and $\rho_{\Delta\mathrm{ML}}$ always has the correct integral with a minimal change in the predicted coefficients.

Because the basis expansion in Equation~(\ref{eq:atom_density}) includes real spherical harmonics of degree $l>0$, which assume both positive and negative values, the raw basis-set density $\rho_{\bomega}(\rv)$ can become locally negative in certain regions of space.
To guarantee that all values of the density are nonnegative, we apply the softplus function to the raw prediction $\rho_{\bomega}(\rv)$, yielding the following function:
\begin{equation}
  \rho_+(\rv) = \mathrm{softplus}_\kappa(\rho_{\bomega}(\rv)) = \frac{1}{\kappa}\log(1+\exp(\kappa\,\rho_{\bomega}(\rv))).
\end{equation}
The softplus function is infinitely differentiable and approximates the identity for positive arguments and zero for negative arguments, with a smooth crossover near $\rho_{\bomega} = 0$. The sharpness of this crossover is controlled by $\kappa$. 
Here, we use $\kappa = 10^8$, corresponding to a crossover density $\rho_c = 1/\kappa = 10^{-8}$~a.u., below which the function deviates appreciably from the identity. 
Note that since the softplus function is not exactly linear for positive arguments, applying it introduces a small normalization error. 
In particular, the integral of $\rho_+(\rv)$ exceeds $N_\mathrm{el}$ by an amount that decreases exponentially with $\kappa$. 
Also note that, in practice, we apply the softplus function only for evaluating metrics involving integrals of the density or its odd powers, e.g., energy functionals. 
When using the softplus function, charge conservation is exact in a grid-based approach, and is approximate with our coefficient-based analytic projection (Equation~(\ref{eq:integral_proj})) with errors that are negligible for $\kappa=10^8$.

Contrary to expectations, our experiments found that applying these constraints during the early training process is not beneficial. The constrained optimization landscape appears to hinder the initial learning of density features, likely because the projection operations interfere with gradient flow during the critical early stages of training. We found that it is most effective to apply these constraints after the model has already been trained and, subsequently, to fine-tune the loss using the constrained model. This two-stage approach allows the network first to learn an accurate unconstrained density representation and then to enforce the physical constraints through a brief fine-tuning phase (see also Bogojeski\cite{bogojeski2023machine} for details).

\subsection{Energy network for molecular dynamics}\label{sec:dens_md}

As shown in previous work,\cite{brockherde2017bypassing, bogojeski2020quantum} the density can serve as a good descriptor for predicting molecular energies and forces. The electron density encodes rich information about the electronic structure relevant to determining the potential energy surface, including bonding patterns, charge distributions, polarization effects, and reactivity. Therefore, we extended the PhiSNet density model by introducing an energy network that takes atomic positions and density model parameters as inputs and outputs the total energy. This modular architecture separates the learning of electronic structure (the density network) from the learning of energetics (the energy network), allowing each component to be optimized independently. The corresponding forces can then be obtained by taking the gradient of the predicted energy with respect to the atomic coordinates, which is simplified by the automatic differentiation libraries built into all modern neural network frameworks.

\paragraph{Energy network architecture.}
The energy network takes the predicted density-fitting parameters and maps them to a total energy using the same message-passing architecture as the density network, thereby inheriting its equivariance properties. The key difference is in how the atom-wise features are initialized: rather than starting from element-based embeddings, the initial representations $\y_i$ are built directly from the atom-wise density basis parameters. The higher-degree spherical components are initialized from the angular parameters $\bgamma^{(l)}$ ($l>0$), while the scalar initial features $\y_i^{(0)}$ are obtained from the radial basis parameters $\bbeta$ and $\balpha$, as well as from the $l=0$ angular parameters $\bgamma^{(0)}$.
These initial features $\y$ and the atomic coordinates $\Rr$ then pass through 
three rounds of message passing, identical in structure to those in the density 
network (see Supplementary Note~1 for the module equation), producing atom-wise representations 
$\ev_i$ that encode how each atom's local density environment contributes to the 
energy. The main distinction between the message passing layers for density and energy prediction is that the energy prediction layers use normalization, since the density model parameters used as input to the energy model can differ widely in scale. Normalization serves to attenuate these differences and makes training more stable (see Supplementary Note~1 for more details on the normalization techniques used).

After the message-passing layers produce the equivariant atom-wise representations, these representations are contracted to rotationally invariant ($l{=}0$) features, to mirror the symmetry of the energy itself, after which a final layer maps each atom's scalar feature to an atomic energy 
contribution $E_i$, and the total energy is obtained by summation $E\ML = \sum_{i}^{M}E_i$. See Supplementary Note~1 for more detailed information about the energy network architecture. 
Forces $\Fv\ML$ follow as the negative gradient of $E\ML$ with respect to the 
atomic positions, computed via automatic differentiation.

To obtain the energies and forces alongside the electron densities, we trained the combined model in two steps. 
First, the density model as described above is trained using the loss function in Equation~(\ref{eq:perc_dens_diff}). 
After the training for the density has converged, the density model parameters are frozen, and the energy network is trained until convergence using the loss function:
\begin{equation}\label{eq:forces_loss}
  \mathcal{L}_{\mathrm{en}} = \frac{1}{N_{\mathrm{batch}}}\sum_{b=1}^{N_{\mathrm{batch}}}\left\{ \lambda_E(|E\ML_b - E_b|) + \lambda_F\left\lVert\Fv\ML_b - \Fv_b\right\rVert_1\right\},
\end{equation}
where $\lambda_E$ and $\lambda_F$ are weights for each term chosen to balance the relative importance of energy and force errors, and $E\ML_b$ and $\Fv\ML_b$ are the predicted energy and forces for the input $b$ in the batch, while $E_b$ and $\Fv_b$ are the corresponding reference energies and forces. See Supplementary Note~2 for a discussion on alternate training strategies.

\subsection{Benchmarking on small organic molecules}\label{sec:benchmark_eval}

Having defined both the density and energy networks, we evaluated their performance on three benchmark molecules, namely ethanol, ethanethiol, and resorcinol. All reference DFT calculations and ML training used the PBE exchange-correlation functional with the cc-pVDZ basis for ethanol and aug-cc-pVDZ for ethanethiol and resorcinol~(Supplementary Note~2). These molecules were chosen as challenging targets for distinct reasons, each representing different chemical environments and bonding characteristics. 
The DenSNet model employed in these experiments uses 128 feature channels, 128 radial basis functions, and a maximum angular degree $L_{\max}$ of 1, 3 and 5 for the first, second and third message passing blocks respectively. The energy function employs a similar but inverted architecture, with 128 feature channels, 32 radial basis functions and a maximum angular degree of 5, 3, 1 for its three message passing layers. Both the energy and density network are similar in size and consist of roughly 3M parameters each, yielding roughly 6M trainable parameters in total.

For each small molecule a separate model was trained on 900--1000 geometries (with 100 held out for validation) for 300\,000 gradient steps, corresponding to approximately 3\,300--4\,500 effective epochs, until convergence; see Fig.~\ref{fig:overview}(b) for the training convergence curve and Supplementary Note~2 for full training details, split sizes, and hardware.

\paragraph{Prediction accuracy.}\label{sec:dens_eval}
To assess how our model captures the ground-state electronic structure, we compare the learned density against other established levels of theory, relative to the reference level of theory. 
As context, for resorcinol, the density absolute fractional error (AFE) for the local-density approximation (LDA) density is $5.4\times10^{-3}$, while Hartree--Fock (HF) densities exhibit an AFE of $2.1\times10^{-2}$. 
Conventional density fitting (DF) basis projections achieve AFE values of $4.9\times10^{-3}$--$6.2\times10^{-3}$. 
These numbers establish the scale of errors between standard quantum-chemical approximations and the chosen Kohn--Sham (KS) PBE reference.
As discussed above, we employed a $\Delta$-learning strategy in which the model is trained on the difference between the true density and the superposition of free-atom densities (SAD). The predicted $\Delta$-learning density is then given by $\rho_{\mathrm{SAD}} + \rho_{\Delta\mathrm{ML}}$, where $\rho_{\Delta\mathrm{ML}}$ represents the learned correction.

\begin{figure}[t]
  \centering
  \includegraphics[width=\linewidth]{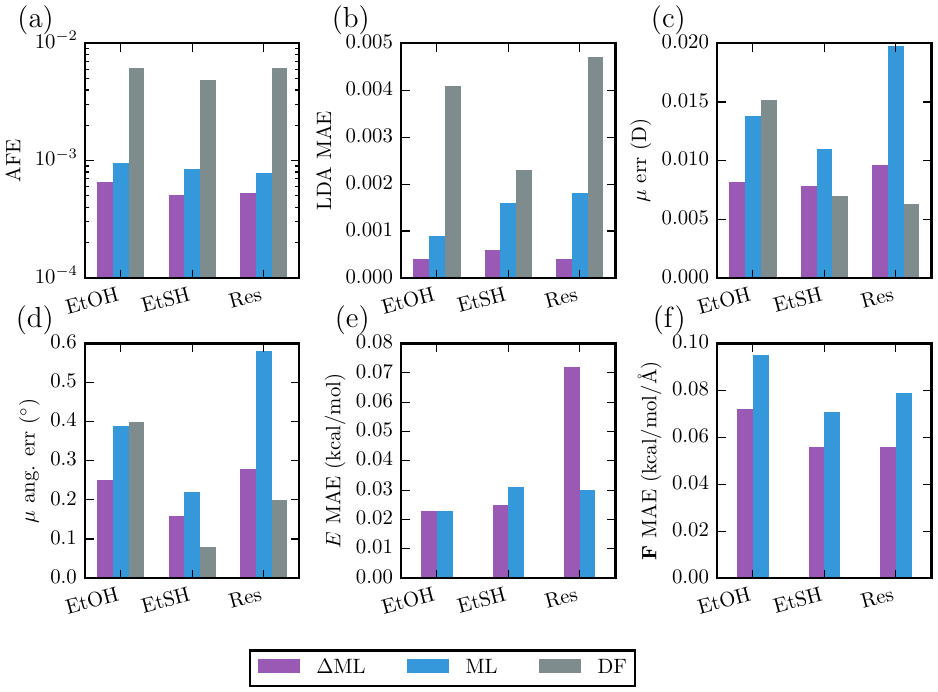}
  \caption{Comparison of density representations across six error metrics for ethanol, ethanethiol, and resorcinol. \textbf{(a)} Density absolute fractional error (AFE). \textbf{(b)} LDA mean absolute error (MAE). \textbf{(c)} Dipole moment magnitude error. \textbf{(d)} Dipole angular error. \textbf{(e)} Energy MAE. \textbf{(f)} Force MAE. Series: $\rho_{\mathrm{SAD}}+\rho_{\Delta\mathrm{ML}}$ ($\Delta$-learning), $\rho_{\mathrm{ML}}$ (direct ML), $\rho_{\mathrm{DF}}$ (density fitting, projection only). Panels (e) and (f) compare energy and force errors for the two ML approaches only; density fitting does not yield a potential or forces. Numerical values are given in Supplementary Table~2.}
  \label{fig:benchmark_results}
\end{figure}

\begin{figure}[t]
  \centering
  \includegraphics[width=\linewidth]{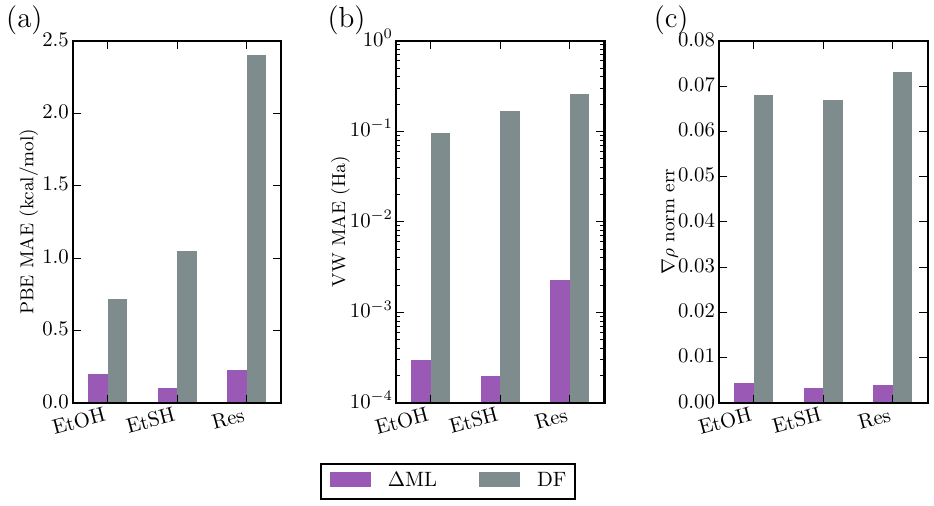}
  \caption{Density-gradient-dependent errors for $\Delta$-learning ($\rho_{\mathrm{SAD}}+\rho_{\Delta\mathrm{ML}}$) versus density fitting ($\rho_{\mathrm{DF}}$) for ethanol, ethanethiol, and resorcinol. \textbf{(a)} PBE exchange-correlation energy MAE (kcal/mol). \textbf{(b)} Von Weizsäcker kinetic energy functional MAE (Ha; log scale). \textbf{(c)} Integrated $\nabla\rho$ norm error. Numerical values are given in Supplementary Table~3.}
  \label{fig:density_grad_err}
\end{figure}

\begin{figure}[!tb]
  \centering
  \includegraphics[width=\linewidth]{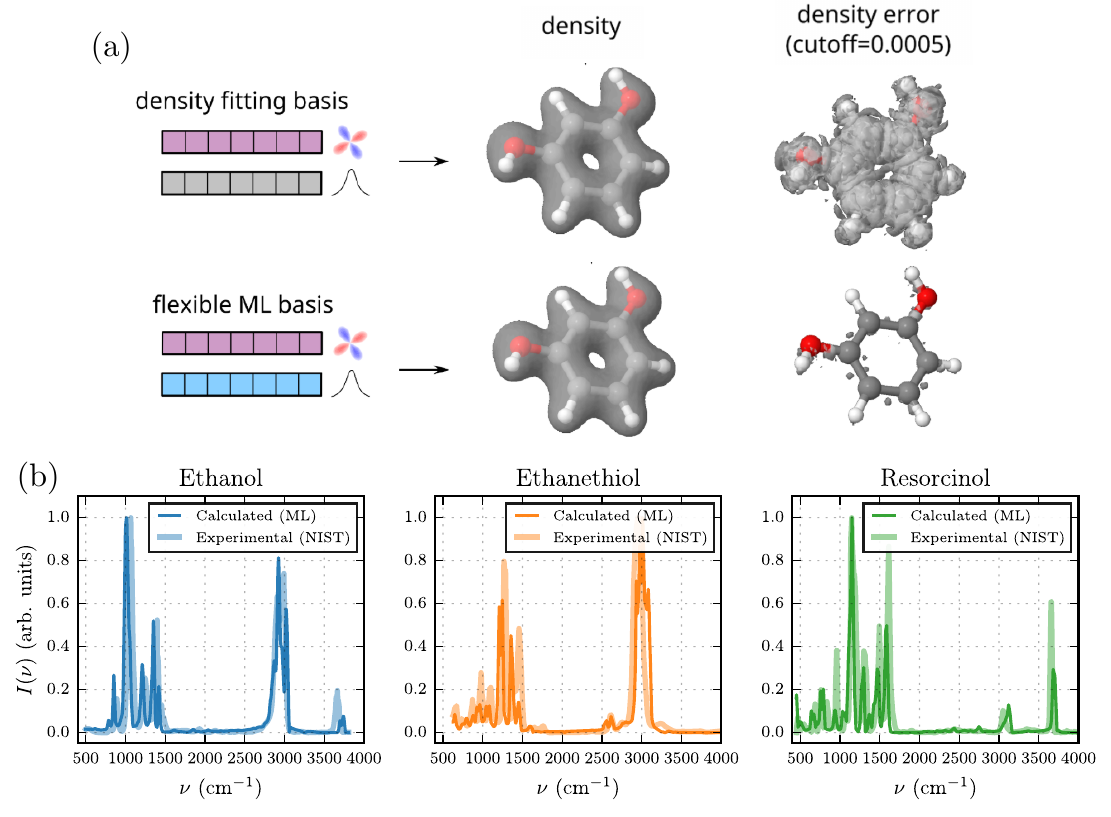}
  \caption{Density prediction accuracy and infrared spectra for small organic molecules. Detailed error metrics are provided in Figure~\ref{fig:benchmark_results}. \textbf{(a)} Comparison of conventional density fitting (DF) and the flexible ML basis for resorcinol. Top row: isosurfaces of the electron density (DF left, ML right). Bottom row: isosurfaces of the density error (cutoff = 0.0005 a.u.). The density fitting basis exhibits substantially larger errors (visible as an extended error cloud), whereas the flexible ML basis achieves markedly lower errors. \textbf{(b)} Infrared absorption spectra for the same three molecules. Dark traces show spectra computed from ML-MD trajectories via the dipole autocorrelation function. Light traces show experimental gas-phase spectra from the NIST database. Peak positions and relative intensities show excellent agreement across all three molecules, validating the accuracy of predicted densities and forces for computing spectroscopic observables.}
  \label{fig:three_molecules}
\end{figure}

Figure~\ref{fig:benchmark_results} presents a comparison of different density representations across multiple error metrics. 
Our $\Delta$-learning approach ($\rho_{\mathrm{SAD}}+\rho_{\Delta\mathrm{ML}}$) achieves AFE values of $5.1\times10^{-4}$--$6.6\times10^{-4}$ across the three molecules, which are roughly 10$\times$ and 40$\times$ more faithful to the PBE reference than LDA ($5.4\times10^{-3}$) and HF ($2.1\times10^{-2}$) calculations, respectively. 
Improvement over conventional density fitting (DF, AFE $4.9\times10^{-3}$--$6.2\times10^{-3}$) is approximately 10$\times$, surpassing DF-basis ground truth used by most prior ML density models.~\citep{brockherde2017bypassing, grisafi2018transferable, fabrizio2019electron, rackers2022cracking, lee2022predicting} This result is expected given that our model is trained directly on KS-DFT density (Supplementary Note~2) rather than fitting DF-basis coefficients, which is the standard in prior works.~\citep{brockherde2017bypassing, grisafi2018transferable, fabrizio2019electron, rackers2022cracking, lee2022predicting}
Note that the improvement is visually apparent in Figure~\ref{fig:three_molecules}(a), where the density error for conventional density fitting is substantially larger than for the flexible ML basis across all three molecules.

The $\Delta$-learning strategy yields an additional 1.5--2$\times$ improvement over learning the full density from scratch (the direct ML model, $\rho_{\mathrm{ML}}$, AFE $7.8\times10^{-4}$--$9.6\times10^{-4}$), demonstrating the advantage of focusing the model's representational capacity on learned corrections.
Related work by Li \textit{et al.}~\cite{li2024superres} also uses a $\Delta$-learning strategy for electron density prediction, refining a coarse density with a CNN on a real-space grid; our approach differs in using an atom-centered basis and coupling to an energy network for dynamics.
The local density approximation (LDA) functional error shows similar trends, with $\Delta$-learning achieving 3--4$\times$ lower error than direct ML. $\Delta$-learning also improves dipole moment predictions, reducing both the magnitude error ($\lVert\mu\rVert$ err) and angular error compared to direct ML, while approaching the accuracy of density fitting in most cases.

The energy and force errors for the ML energy network (panels e--f of Figure~\ref{fig:benchmark_results}) are well below the standard threshold of 1 kcal/mol(/\AA) for both delta-learned densities ($\rho_{\mathrm{SAD}}+\rho_{\Delta\mathrm{ML}}$) and direct ML densities ($\rho_{\mathrm{ML}}$). 
The mean absolute errors (MAEs) for all three molecules are lower than the quantum chemical accuracy threshold by at least a factor of ten, in line with other state-of-the-art ML potentials (see Supplementary Note~3).
$\Delta$-learning can exhibit somewhat higher energy MAE than direct ML in some cases, since the $\Delta$-learning energy network is trained with greater weight on forces (Equation~(\ref{eq:forces_loss})) for molecular dynamics; retraining with a higher energy weight would reduce energy MAE if required.

Beyond the integrated density, accurate reproduction of GGA-level exchange-correlation functionals, which are what we used for training the model, requires faithful representation of density gradients.
Figure~\ref{fig:density_grad_err} compares density gradient errors between $\Delta$-learning and density fitting. 
The $\Delta$-learning model achieves substantially lower errors across all metrics: the Perdew--Burke--Ernzerhof (PBE) exchange-correlation energy error is 3--10 times lower (0.1--0.23 vs. 0.72--2.4 kcal/mol), the von Weizsäcker kinetic energy~functional error shows larger improvements (2--3 orders of magnitude lower), and the gradient norm error is 15--20$\times$ lower. 
These results confirm that the flexible basis representation captures not only the integrated density but also its spatial structure. 
In other words, the model is accurate enough to evaluate GGA-level density functionals, validating that it correctly describes the electron distribution at the level required for chemical bonding and reactivity.

\paragraph{Infrared spectra via density enhanced molecular dynamics.}\label{sec:dipole_md}
Accurate energies and forces on benchmark data, while necessary, are not sufficient to guarantee stable behavior in long molecular dynamics simulations,~\cite{chmiela2018towards,fu2022forces} where the system can visit regions of phase space under-represented in the training set. 
To test whether the predicted densities and forces remain accurate over long trajectories, we computed IR spectra for ethanol, ethanethiol, and resorcinol via density-enhanced molecular dynamics.
Since obtaining a converged IR spectrum requires hundreds of thousands to millions of time steps, standard Kohn-Sham (KS) DFT calculations are prohibitively expensive. Our machine learning model produces accurate forces and densities at a cost that is orders of magnitude lower than KS-DFT, making such simulations practical.~\cite{ji2025universal}
The molecular dynamics simulations for obtaining the IR spectra were performed using SchNetPack,~\cite{schutt2018schnetpack} with an NVE ensemble at 300~K and a time step of 0.5~fs. This timestep was chosen to sample the fastest vibrational modes accurately (C-H stretches around 3000 cm$^{-1}$) while maintaining numerical stability. Each trajectory was propagated for sufficient duration to obtain converged spectral estimates, with the dipole moment computed directly from the ML-predicted electron density at every timestep (see Supplementary Note~2 for dataset details and simulation parameters).
The resulting machine-learned IR spectra (Fig.~\ref{fig:three_molecules}b) show excellent agreement with experimental IR spectra recorded in the gas phase at 300~K taken from the NIST database.~\cite{ralchenko2022nist} 
To assess whether the model correctly captures the electronic structure underlying each vibrational mode, we examine the chemical significance of the dominant spectral features for each molecule:

For ethanol, the ML-predicted spectrum closely reproduces the experimental gas-phase spectrum across the full 500--4000~cm$^{-1}$ range.
The free O-H stretching mode of the isolated hydroxyl group appears at $\sim$3650~cm$^{-1}$ in the gas phase, the spectroscopic signature of the alcohol functional group, and is well captured by the model.
Minor differences in peak width for the O-H stretch arise from rotational fine structure and intramolecular vibrational energy redistribution (IVR), quantum-mechanical effects that are not captured in our single-molecule classical MD trajectory at 300~K.~\cite{keshavamurthy2014scaling}
The C-H stretching modes near 3000~cm$^{-1}$, arising from the methyl (CH$_3$) and methylene (CH$_2$) groups, are reproduced with accurate peak positions and intensities.
The fingerprint region below 1500~cm$^{-1}$, which encodes C-O stretching (near 1050~cm$^{-1}$) and skeletal C-C-O deformation modes characteristic of the primary alcohol backbone,~\cite{nistwebbook} shows close agreement in both peak positions and relative intensities, confirming that the learned density and forces correctly describe the bonding environment of the hydroxyl-substituted alkyl chain.

For ethanethiol, the ML-predicted and experimental spectra are in excellent agreement across the full spectral range, with peak positions and relative intensities accurately reproduced for all major vibrational features.
The S-H stretching mode at $\sim$2550~cm$^{-1}$ is the spectroscopic signature distinguishing thiols from alcohols, appearing at substantially lower frequency than the O-H stretch ($\sim$3650~cm$^{-1}$) due to the lower force constant of the S-H bond.
The S-H stretch has an intrinsically weaker IR absorption intensity than the O-H stretch---roughly five times weaker in the fundamental region---caused by weaker dipole moment derivative ($\partial\mu/\partial Q$) for the less polar S-H bond.~\cite{kuodis2020reflection}
The model reproduces both the position and relative intensity of this mode, as well as the C-H stretching cluster near 3000~cm$^{-1}$ and the fingerprint features below 1500~cm$^{-1}$, demonstrating that the learned density correctly captures the distinct electronic environment of sulfur-hydrogen bonding and validates the model's ability to generalize across different heteroatom bonding environments (oxygen vs.\ sulfur).

For resorcinol, the ML-predicted spectrum shows excellent agreement with experiment, accurately reproducing peak positions and relative intensities across the full spectral range.
The prominent mode at $\sim$750~cm$^{-1}$ corresponds to aromatic C-H out-of-plane bending, which is diagnostic of the 1,3-(meta)-disubstitution pattern of the benzene ring.~\cite{chemes2021assessment} Different positional isomers of dihydroxybenzene produce distinct out-of-plane bending signatures, so capturing this mode correctly tests the model's sensitivity to aromatic substitution patterns.
The model reproduces this mode with accurate position and intensity, along with the phenolic O-H stretch at $\sim$3650~cm$^{-1}$ and the C-H stretching modes near 3000~cm$^{-1}$.
The fingerprint region below 1500~cm$^{-1}$, which encodes aromatic ring deformation and C-O stretching modes, is also well captured.
Minor differences in the relative intensities of individual peaks are consistent with the conformational complexity of resorcinol, which has multiple low-energy conformers related to O-H rotational isomerism.~\cite{pitsevich2023torsional}
The close agreement across all spectral regions demonstrates that the model successfully describes an aromatic $\pi$-system with electron-donating substituents, a qualitatively different electronic environment from the aliphatic molecules.

The agreement between ML-predicted and experimental IR spectra across all three molecules, spanning aliphatic alcohols, thiols, and aromatic diols, confirms that the model predicts accurate densities and forces that correctly capture the electronic structure underlying the vibrational modes of chemically distinct functional groups.

\subsection{Extrapolation to larger systems: Polythiophene scaling}

Having validated DenSNet on small organic molecules with diverse bonding environments, we next assess its ability to extrapolate beyond the training distribution. Transferability to larger systems is essential for practical applications in materials science and soft matter, where target systems often exceed the sizes used for training.

\begin{figure}[t]
\centering
\includegraphics[width=\linewidth]{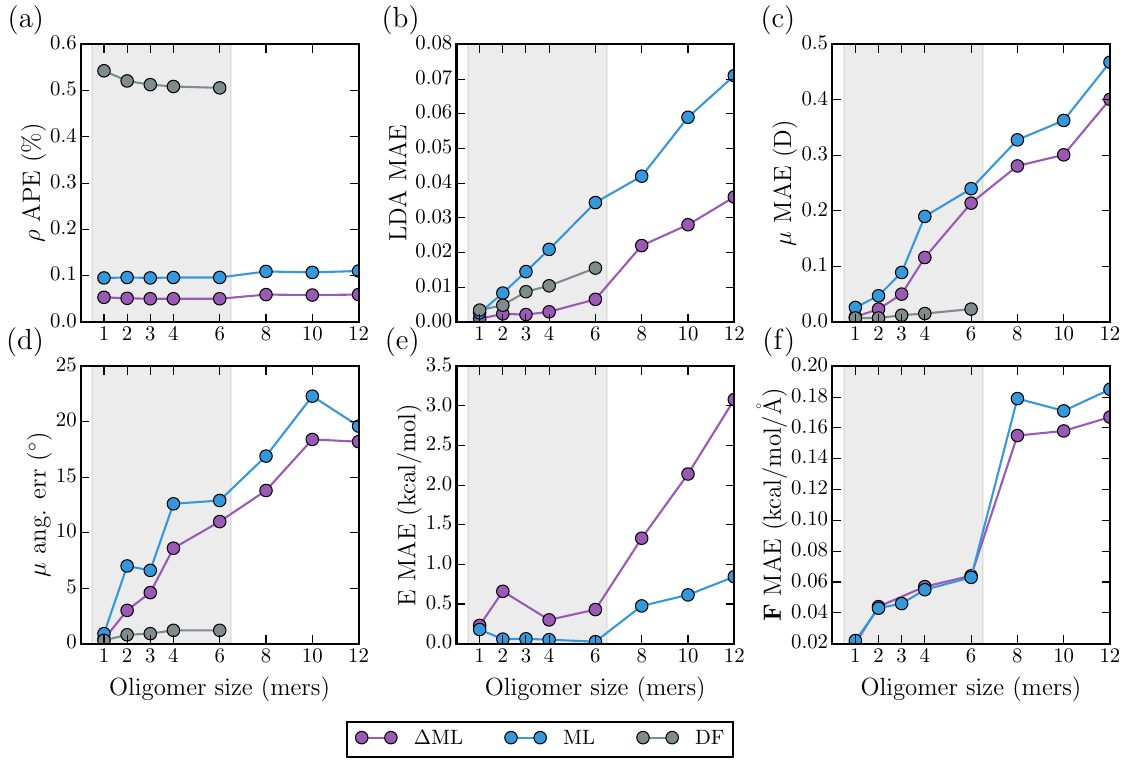}
\caption{Error metrics versus oligomer size for $\rho_{\Delta\mathrm{ML}}$, $\rho_{\mathrm{ML}}$, and $\rho_{\mathrm{DF}}$ across training (1--6 mers) and extrapolation (8--12 mers). Gray shaded region: training regime. \textbf{(a)} $\rho$ AFE. \textbf{(b)} LDA MAE. \textbf{(c)} $\mu$ MAE (Debye). \textbf{(d)} $\mu$ angular error (deg). \textbf{(e)} E MAE (kcal/mol). \textbf{(f)} F MAE (kcal/mol/\AA). Energy and force are shown for $\rho_{\mathrm{ML}}$ and $\rho_{\Delta\mathrm{ML}}$; $\rho_{\mathrm{DF}}$ is a density projection and does not provide energies or forces. Numerical values are given in Supplementary Table~7.}
\label{fig:thiophene_metrics}
\end{figure}

\paragraph{Polythiophene as a benchmark for size transferability.}\label{sec:polythiophene}
To evaluate the size-transferability of DenSNet, we applied it to polythiophene oligomers of increasing size, training on small chains (1--6 monomers) and testing on larger systems (8--12 monomers). 
Thiophene is a building block for organic solar cells, valued for its electronic and redox properties and supramolecular behavior.~\cite{zhang2011thiophene} 
Its conjugated $\pi$-system, with sulfur heteroatoms enhancing electron delocalization, gives rise to cooperative electronic effects as the chain length increases, including bandgap narrowing and enhanced polarizability.~\cite{monkman2001conjugation}
These size-dependent electronic properties provide a stringent test of whether the learned density generalizes to electronic environments not seen during training. Reference DFT calculations and ML training used the PBE functional with the aug-cc-pVDZ basis~(Supplementary Note~4). 

\paragraph{Prediction accuracy for polythiophene oligomers.}

We created a dataset comprising 1000 samples each of 1-, 2-, 3-, 4-, and 6-mer oligomers (5000 samples total), trained on 4500 geometries, and validated on the remaining 500.
The $\Delta$-learning model used a 5~\AA\ cutoff radius to improve generalization to larger polymers.
The model's performance was evaluated on a separate test set sampled from a classical MD trajectory at 300~K.
Our ML-predicted densities achieve approximately 5.5$\times$ lower AFE ($9.5\times10^{-4}$--$9.6\times10^{-4}$) than standard density fitting methods ($5.1\times10^{-3}$--$5.4\times10^{-3}$) across the training regime, confirming that the equivariant learning framework reproduces the PBE electronic structure more faithfully than conventional density-fitting projections. 
$\Delta$-learning ($\rho_{\Delta\mathrm{ML}}$) achieves consistently lower density AFE ($5.0\times10^{-4}$--$5.9\times10^{-4}$) than direct ML ($9.5\times10^{-4}$--$1.1\times10^{-3}$) across all oligomer sizes, roughly halving the density error.

To test the model's ability to extrapolate beyond its training regime, we ran four independent 100~ps ML-MD trajectories for 8-, 10-, and 12-mer oligomers, with systems up to twice the polymer size of the training data. 
The results in Figure~\ref{fig:thiophene_metrics} demonstrate stable extrapolation: the density AFE remains essentially constant across all system sizes (${\sim}1.1\times10^{-3}$), showing no deterioration with increasing molecular complexity; force errors likewise show minimal size dependence (0.171--0.185 kcal/mol/Å), remaining well below the 1 kcal/mol/Å threshold and producing stable MD trajectories throughout the 100~ps simulations.

For dipole moment predictions, DenSNet achieves low MAE (0.014-0.109~D) within the training regime, with angular errors of $7$--$17^\circ$. 
Note that DF achieves substantially lower errors (0.0038--0.0108~D) with angular errors $<1.3^\circ$. 
More importantly, the dipole moment MAE increases with system size (0.146 to 0.211~D in the extrapolation regime, with angular errors of 17--22°). 
Despite this growth of errors, the increase of MAE with system size is a predictable geometric consequence rather than a sign of degrading model quality. 
In polythiophene, the thiophene monomers are oriented with alternating head-to-tail stacking, so their dipole vectors add with effectively random relative phases; by the same argument as a random walk, both the true dipole magnitude and the ML prediction error scale as $a\sqrt{n}+b$. Fitting this model for $n>2$ yields $R^2>0.97$ for both density representations, and the rescaled residual $\varepsilon(n)=(\mathrm{MAE}(n)-b)/\sqrt{n}$ is flat across the entire size range tested, confirming that the \emph{per-monomer prediction error is constant} and does not accumulate with chain length (Supplementary Note~4). 

\begin{figure}[!htb]
  \centering
  \includegraphics[width=0.975\linewidth]{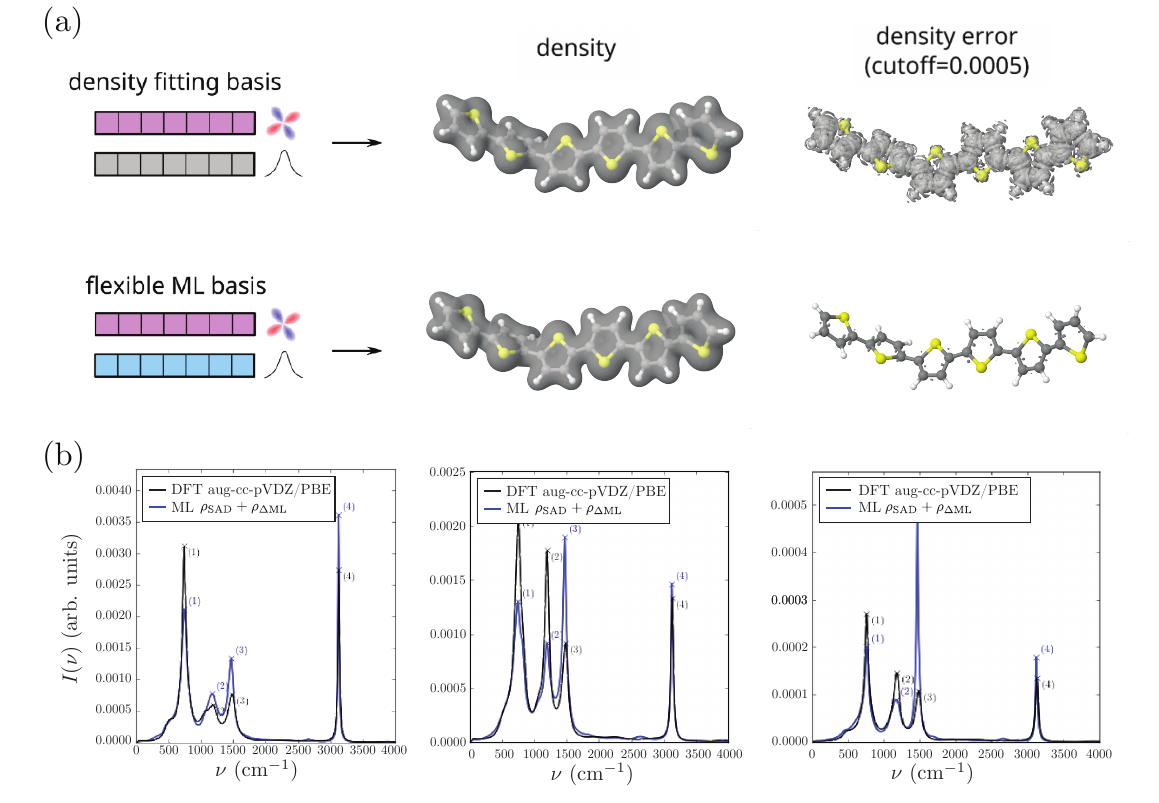}
  \caption{Transferability of DenSNet to polythiophene oligomers. \textbf{(a)} Comparison of conventional density fitting (DF) and the flexible ML basis for polythiophene. Top row: electron density isosurfaces (DFT left, ML right). Bottom row: density error isosurfaces. \textbf{(b)} IR spectra for extrapolation systems (8-, 10-, and 12-mers) comparing ML-MD predictions to DFT reference calculations. Peak positions show accuracy within experimental resolution limits, validating the transferability enabled by SE(3)-equivariant atom-centered representations. Quantitative peak error analysis is provided in Figure~\ref{fig:ir_peak_errors}.}
  \label{fig:thiophene_scaling}
\end{figure}

\paragraph{IR spectroscopic evaluations.}
To subject our model to the most rigorous test of extrapolation, we used the best-performing $\Delta$-learning model for IR spectroscopic evaluation of the 8-, 10-, and 12-mer systems.
We assessed spectroscopic accuracy by computing IR spectra from the dipole moment time series using the dipole autocorrelation function (DACF).
Four 100~ps ML-MD trajectories were initiated from geometries constructed by concatenating smaller oligomer units. 
ML dipole moments were calculated directly from DenSNet, while reference DFT dipole moments were computed using PySCF (Kohn-Sham DFT, PBE/aug-cc-pVDZ). 
The DACF was computed via direct time-averaging, and the corresponding spectra were obtained using the maximum entropy method with Final Prediction Error (MaxEnt-FPE). 
The MaxEnt approach is chosen over an FFT approach due to the limited number of samples and short trajectories we obtained, and FPE is used to fix the hyperparameters of the MaxEnt approach.

The resulting IR spectra (Fig.~\ref{fig:thiophene_scaling}b) demonstrate close agreement between ML-MD predictions and DFT reference calculations for the extrapolation systems.
To assess whether the model correctly captures the electronic structure of the conjugated backbone, we examine the chemical significance of the dominant vibrational modes:
The C=C antisymmetric ring stretching mode near $\sim$1440~cm$^{-1}$ is the most prominent in-plane vibration of the thiophene ring and is sensitive to the degree of conjugation along the polymer backbone;~\cite{navarrete1991lattice} its frequency and intensity encode information about backbone planarity and inter-ring torsional disorder, providing a direct probe of the $\pi$-electron delocalization that governs the optoelectronic properties of polythiophene materials.
The C-H out-of-plane bending mode near $\sim$790~cm$^{-1}$ is characteristic of the thiophene heterocycle and is a well-established diagnostic signature in oligothiophene and polythiophene spectroscopy.~\cite{navarrete1991lattice,parker2023vibrational}
Reproducing these modes validates that the ML forces capture the local curvature of the potential energy surface around the equilibrium geometry of the conjugated backbone.
Note that the spectral convergence from 8- to 12-mers also mirrors known experimental and computational behavior, where oligothiophene vibrational spectra converge toward the polymer limit by approximately the hexamer level.~\cite{parker2023vibrational}

\begin{figure}[t]
\centering
\includegraphics[width=\linewidth]{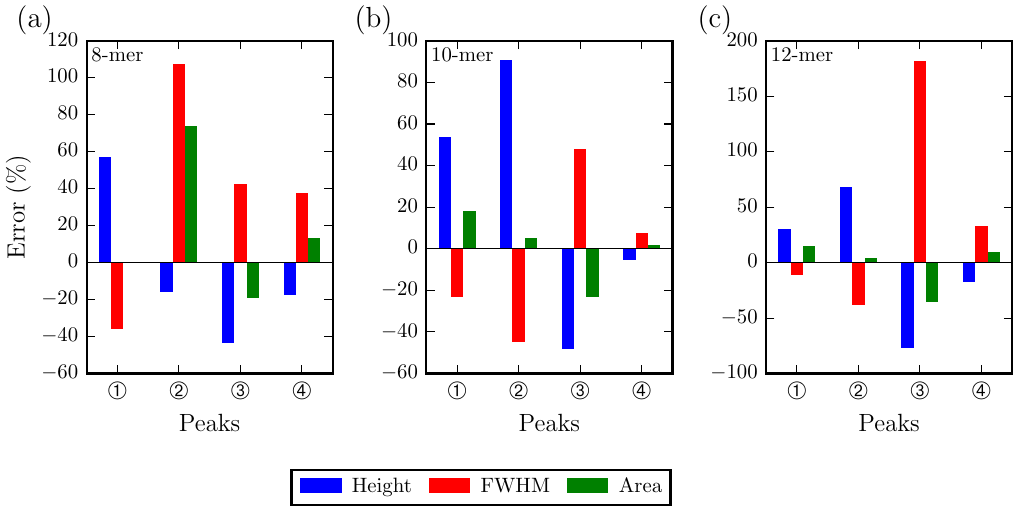}
\caption{Quantitative IR peak errors for polythiophene 8-, 10-, and 12-mers. \textbf{(a)} 8-mer. \textbf{(b)} 10-mer. \textbf{(c)} 12-mer. Percentage error in peak height, FWHM, and area for the four dominant peaks (averaged over four 100~ps trajectories). Peak heights and FWHM show anti-correlated errors; integrated areas are the physically meaningful intensity metric and remain within 4--18\% for most peaks. Numerical values are given in Supplementary Table~6.}
\label{fig:ir_peak_errors}
\end{figure}

Quantitative peak analysis required careful spectral fitting due to overlapping peak structures in the oligomers.
To this end, we used a custom fitting routine using Voigt profiles (Gaussian-Lorentzian convolutions) with manual peak initialization, extracting four key metrics, namely peak position, amplitude, full-width at half-maximum (FWHM), and integrated area.
Peak positions showed consistent accuracy across all system sizes ($<1\%$ error, typically 3--15~cm$^\mathrm{-1}$), well within experimental IR resolution limits ($\pm$2~cm$^\mathrm{-1}$).
Intensity metrics (amplitude, FWHM) exhibited anti-correlated errors, as expected for finite sampling.
Integrated peak areas, which are proportional to the square of the transition dipole moment derivative ($\partial\mu/\partial Q$),  directly reflect the accuracy of the predicted density's response to nuclear displacements; all of which demonstrated good agreement with DFT reference values as seen in \cref{fig:ir_peak_errors}.
For the four dominant peaks across 8-, 10-, and 12-mer systems, area errors ranged from 4--36\% for most peaks, though some weak, broad features showed larger discrepancies (up to 74\%, with detailed quantitative analysis in Figure~\ref{fig:ir_peak_errors}).
Individual trajectories occasionally exhibited anomalously intense peaks (e.g., 12-mer trajectory 1), highlighting the importance of ensemble averaging for converged spectroscopic properties.

The systematic accuracy in peak positions, combined with reasonable agreement in integrated intensities, validates the model's ability to predict vibrational spectroscopic observables directly from ML-MD trajectories without ab initio calculations at every timestep.
This enables spectroscopic characterization of conjugated polymer conformational dynamics over timescales inaccessible to conventional DFT-MD, though care must be taken to ensure adequate conformational sampling through multiple independent trajectories.
The close agreement for chemically diagnostic modes (C=C ring stretch, C-H out-of-plane bending) demonstrates that the learned density and forces correctly capture the electronic structure of the conjugated thiophene backbone at the PBE level of theory.

Figure~\ref{fig:ir_peak_errors} provides a detailed quantitative analysis of the errors in predicted IR spectral features across the three extrapolation systems, where several key observations can be made. 
First, peak positions exhibit high accuracy, with errors consistently below 5\% (corresponding to 3--15~cm$^{-1}$ absolute differences), well within the experimental IR resolution limit of $\pm$2~cm$^{-1}$.
This systematic accuracy in vibrational frequencies demonstrates that the ML-predicted forces correctly capture the local curvature of the potential energy surface.
Second, peak heights and full-widths at half-maximum (FWHM) show substantially larger errors, ranging from 6--91\% for heights and 8--182\% for widths, but anticorrelate with one another, i.e.,  when the peak height is underestimated, the FWHM tends to be overestimated, and vice versa.
This anti-correlation motivates us to calculate the integrated peak areas, which are proportional to the transition dipole strength $|\partial\mu/\partial Q|^2$ and thus directly measure how accurately the learned electronic structure responds to nuclear motion. 
With this metric, we see good agreement with DFT reference values;
For the majority of peaks across all three oligomer sizes, area errors remain in the 4--18\% range, with only one notable outlier (8-mer Peak 2 at 74\% error, which corresponds to a weak, broad feature with significant overlap).
This consistency in integrated intensities validates the ML model's ability to correctly predict both the magnitudes of the transition dipole moments and their time-dependent fluctuations, the fundamental quantities underlying IR absorption.

\FloatBarrier
\section{Discussion}

We have presented DenSNet, a density-equivariant deep learning framework that enables molecular dynamics simulations with direct access to electronic properties beyond energies and forces. Our approach combines SE(3)-equivariant neural networks with $\Delta$-learning from superposed atomic densities and a modular energy network. This strategy yields sub-chemical-accuracy force predictions (MAE $<$ 0.1 kcal/mol/Å) while simultaneously providing accurate electron densities (AFE ${\sim}10^{-3}$) that enable direct computation of infrared spectra, dipole moments, and other electronic observables during MD trajectories. The excellent agreement between our machine-learned IR spectra and experimental measurements for ethanol, ethanethiol, and resorcinol demonstrates the reliability of the predicted densities and forces for computing macroscopic observables. The successful extrapolation to polythiophene oligomers twice as large as the training systems (12-mers vs 6-mers) validates the transferability of DenSNet to larger system sizes.
By learning the density rather than a collection of disconnected models for the observables, DenSNet reinstates the electron density as the central object in molecular dynamics, unifying dynamics and electronic observables within a single, physically grounded representation.

DenSNet can be immediately applied to vibrational spectroscopy of extended $\pi$-conjugated systems, such as conjugated polymers, donor--acceptor copolymers, and organic photovoltaic materials, where MLIPs alone cannot provide dipole derivatives for IR intensities.~\cite{mai2025computing,bhatia2025leveraging} For solvated molecules and hydrogen-bond networks, DenSNet also enables computation of dipole autocorrelation functions and frequency-dependent dielectric response from large-scale trajectories.~\cite{krishnamoorthy2021dielectric} 
Finally, density-based observables along reaction coordinates, such as charge redistribution during proton-coupled electron transfer, can now become accessible through DenSNet without expensive on-the-fly electronic structure calculations.~\cite{hammesschiffer2008proton} Although DenSNet is slower than AIMNet2 due to its use of higher-order equivariant features, it is already orders of magnitude faster than \textit{ab initio} MD at the 12-mer scale (Supplementary Note~4) and, unlike AIMNet2 whose dipole moments are approximated by fixed point charges, it provides full access to all density-derived electronic observables on the fly.
The predicted density also has two direct uses: as a Kohn--Sham initial guess, it can reduce SCF iterations to one or two and enable direct XC energy checks during an MD trajectory; and it applies to any existing MLIP, providing density-derived observables without retraining.

Incorporating SE(3) equivariance into the model architecture has practical advantages: it reduces data requirements by eliminating the need to learn identical physics under different orientations, and learning the physical electron density rather than abstract latent representations provides interpretability alongside accuracy. 
The stability of our MD simulations over hundreds of picoseconds further confirms the reliability of the learned force fields.
Equivariant MLIPs such as MACE,~\cite{batatia2022mace} NequIP,~\cite{batzner2021nequip} and SO3krates~\cite{frank2022so3krates,frank2024euclidean,kabylda2025so3lr} have improved energy and force accuracy. 
Recent extensions also target IR spectroscopy, such as MACE4IRmol~\cite{bhatia2025mace4ir} which trains a separate dipole model alongside the potential, while MACE-POLAR-1~\cite{batatia2026macepolar} incorporates learnable charge densities via a polarizable electrostatic formalism. In both cases, however, the dipole is a separately learned quantity rather than a consequence of the full electron density, limiting access to other electronic observables. Other approaches learn atom-centered densities but stop at a single Kohn--Sham diagonalization,~\cite{grisafi2022electronic,rackers2022cracking} predict Hamiltonians without coupling to dynamics,~\cite{kaniselvan2025helm} or adapt density functionals without bypassing the Kohn--Sham equations.~\cite{khan2025adapting} 
Our work, which builds upon the ML--Hohenberg--Kohn map~\cite{brockherde2017bypassing} and density-based $\Delta$-learning,\cite{bogojeski2020quantum,shao2023machine} establish the key tenet that the electron density serves as a universal descriptor for both electronic observables and molecular dynamics. In particular, DenSNet extends this program with SE(3) equivariance and an atom-centered Gaussian basis coupled to a modular energy network, yielding an end-to-end pipeline from nuclear coordinates to electron density to molecular dynamics to IR spectra, with demonstrated transferability to systems twice the training size.

The framework presented here is not limited to ground-state densities. Recent work has demonstrated that machine learning can be extended to predict excited-state electron densities.~\cite{bai2022machine} This enables simulations of photochemical processes and non-adiabatic dynamics without the computational burden of time-dependent DFT at every timestep. Learning the response of the density to electric fields (e.g., for polarizability) has been addressed in a related vein.~\cite{lewis2025electrostatic} Extending the approach from scalar densities $\rho(\rv)$ to full one-particle density matrices $\gamma(\rv, \rv')$~\cite{shao2023machine} would give access to natural orbitals and improved treatment of strong correlation while preserving the efficiency of machine learning.

Large-scale, curated open datasets such as the Open Catalyst Project,~\cite{chanussot2021open} Open Molecules,~\cite{levine2025open} and QCML,~\cite{ganscha2025qcml} now include energies, forces, and electronic structure data spanning millions of configurations. DenSNet can exploit such datasets because electron density is a universal descriptor of electronic structure: models pre-trained on large, heterogeneous datasets can be fine-tuned for specialized applications with minimal additional data. Combining physics-informed architectures (equivariance, $\Delta$-learning) with data-driven pre-training should enable few-shot transfer to previously unseen chemical systems.

Lastly, while DenSNet successfully predicts IR spectra and stable MD trajectories, the density absolute fractional error (AFE) does not directly optimize dipole moments, which depend on the first spatial moment of the density rather than its integrated magnitude. As a result, dipole moment mean absolute errors are higher than those obtained from conventional density fitting, particularly for larger oligomers. Minimizing the density AFE loss function does not directly translate into improved molecular dipole-moment accuracy (see Supplementary Note~3 for detailed dipole moment error analysis).~\cite{wayo2025qdftnet} This discrepancy indicates that a loss function explicitly incorporating dipole constraints could yield better predictions of electrostatic properties. Importantly, however, the raw MAE growth with oligomer size is not a sign of deteriorating model quality: the per-monomer error is constant across all tested sizes, and the observed increase follows the expected $\sqrt{n}$ random-walk scaling of the dipole moment itself (Supplementary Note~4). Nonetheless, IR integrated intensities remain accurate, demonstrating that DenSNet correctly captures the time-dependent dipole fluctuations essential for vibrational spectroscopy. Future work incorporating dipole-regularized loss functions or multi-task learning objectives may reconcile density accuracy with improved electrostatic property prediction, while preserving the unified, density-first approach central to this framework.

\FloatBarrier
\section{Methods}\label{sec:methods}
In this section, we provide a concise overview of the reference electronic structure calculations, the molecular dynamics setup, and the equivariant neural network model. Detailed settings for basis sets, grids, training procedures, and extended benchmarks are collected in the SI. Note that a single DenSNet trajectory step costs approximately 755~ms on a single GPU for polythiophene (86~atoms), compared to ${\sim}$5~min for a Kohn--Sham DFT step (extrapolated fit to GPU-accelerated PySCF benchmarks), making density-enhanced MD through DenSNet roughly 400$\times$ cheaper than AIMD while providing full electronic structure at every timestep (Supplementary Note~4).

\subsection{Error measures}\label{sec:error_measures}
Throughout this work, we used several error metrics to quantify the accuracy of predicted electron densities and derived properties. Let $M$ denote the number of samples, $\tilde{\rho}$ the expected density, $\rho$ the reference density, and $\mu[\rho, \mathbf{R}, Z]$ the dipole moment calculated from a density $\rho$ with nuclear positions $\mathbf{R}$ and atomic charges $Z$. The primary error measures are:
\begin{equation*}
\begin{aligned}
\rho\text{ AFE} &= \frac{1}{M}\sum_{m=1}^{M}\frac{\int \left|\tilde{\rho}_{m}(\rv) - \rho_m(\rv) \right| d\rv}{\int \rho_m(\rv) d\rv}\\
\text{LDA MAE} &= \frac{1}{M}\sum_{m=1}^{M}\left| \int \tilde{\rho}_{m}(\rv)^{4/3}d\rv - \int \rho_m(\rv)^{4/3}d\rv\right| \\
\mu\text{ err} &= \frac{1}{M}\sum_{m=1}^{M}\lVert \mu[\tilde{\rho}_{m}] - \mu[\rho_m] \rVert\\
\lVert\mu\rVert\text{ err} &= \frac{1}{M}\sum_{m=1}^{M}\left| \lVert\mu[\tilde{\rho}_{m}]\rVert - \lVert \mu[\rho_m]\rVert \right|\\
\mu \text{ ang err} &= \frac{1}{M}\sum_{m=1}^{M} \arccos{\left(\frac{\mu[\tilde{\rho}_{m}]\cdot\mu[\rho_{m}]}{\lVert\mu[\tilde{\rho}_{m}]\rVert \lVert\mu[\rho_{m}]\rVert}\right)}
\end{aligned}
\end{equation*}
The AFE (absolute fractional error) measures the overall integrated density error normalized by the total electron count. The LDA MAE captures errors in the local-density-approximation energy functional. The dipole moment error ($\mu$ err) measures the Euclidean distance between predicted and reference dipole vectors, while $\lVert\mu\rVert$ err and $\mu$ ang err decompose this into magnitude and angular components, respectively.

\subsection{Reference DFT calculations}\label{sec:ref_calc}
All reference electronic structure calculations were performed using PySCF with DFT.
For ethanol, PBE/cc-pVDZ was used. For ethanethiol and resorcinol, PBE/aug-cc-pVDZ was employed.
Electron densities were evaluated directly from the wavefunction on angular-radial integration grids (see Supplementary Note~2 for grid resolution analysis).
Dataset generation details, including conformer sampling and $k$-means selection procedures, are provided in Supplementary Note~2.

\subsection{Molecular dynamics}\label{sec:md}
ML-MD simulations were performed using SchNetPack~\cite{schutt2018schnetpack} with an NVE ensemble at 300~K and a timestep of 0.5~fs.
IR spectra were computed from the dipole autocorrelation function using the Maximum Entropy method with Final Prediction Error (FPE) regularization.
For polythiophene oligomers, four independent 100~ps trajectories were initiated from geometries constructed by concatenating smaller oligomer units.
Extended trajectory analysis and per-trajectory results are provided in Supplementary Note~4.

\begin{figure}[t]
  \centering
	\includegraphics[width=\textwidth]{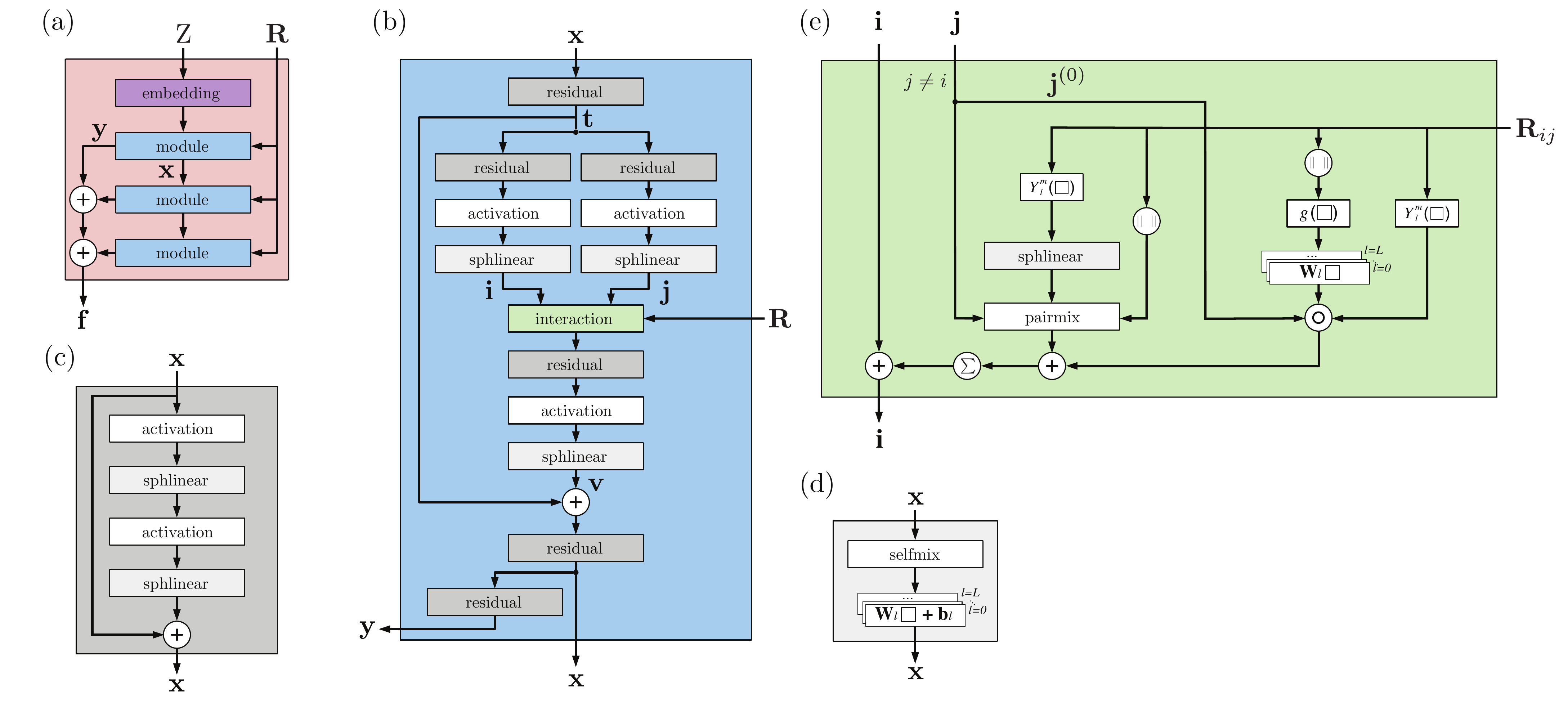}
  \caption{\footnotesize Illustration of the representation module components, based on PhiSNet. \textbf{(a)}: Generation of atomic spherical harmonics features $\mathbf{f}$. The Cartesian coordinates $\{\mathbf{r}_i\}$ of the atoms are used to calculate a spherical harmonics representation of the relative atomic positions. An embedding layer (purple) creates initial atomic features from the nuclear charges $\{Z_i\}$, which are refined through a series of equivariant modular blocks (blue). The outputs $\mathbf{y}$ of each modular block are summed to obtain the final representations $\mathbf{f}$. \textbf{(b)}: Modular block. Input representations are separated into two branches, which produce separate features for the central $\mathbf{i}$ and neighboring atoms $\mathbf{j}$. These are coupled by an interaction block (green) and added to the original features to produce the $\mathbf{x}$ and $\mathbf{y}$ outputs. \textbf{(c)}: Residual blocks
pass input features through two non-linear activations and spherical linear layers, and add the result to the unmodified input via a skip-connection. \textbf{(d)}: Spherical linear layers
are composed of a self-mix layer followed by separate linear layers for each spherical harmonic order, and are used to re-combine spherical harmonic orders and feature channels. \textbf{(e)}: The interaction block encodes information about chemical environments by combining features of neighboring atoms with a spherical harmonics-based representation of their relative position to a central atom.} 
\label{fig:nn_architecture}
\end{figure}

\subsection{Equivariant neural network}\label{sec:equiv_net}
Deep message-passing neural networks (MPNNs).~\citep{gilmer2017neural} for quantum-chemistry applications model the physical properties of atomistic systems as an aggregate of atomic contributions, which are predicted from representations of each atom's local environment, $\mathbf{x}_i \in \mathbb{R}^C$. Using element-specific embeddings as the starting point, the atomic representations are constructed by iteratively exchanging messages between each atom $i$ and its neighbors $j$. These messages depend on the current atomic representations of the center atom $\mathbf{x}_i$, its neighbors $\mathbf{x}_j$, and respective distances $r_{ij}$, and are used to update the state of $\mathbf{x}_i$ after each interaction.

If geometric information about the molecule is provided only as pairwise distances, the final atomic features will be rotationally invariant by construction, which is desirable when they are used to predict a quantity that itself is rotationally invariant, for example, the potential energy. Recent developments in message passing architectures~\citep{batzner2021nequip,schutt2021equivariant, unke2021spookynet} show that including directional information about the atom-pair both increases the prediction accuracies for invariant properties and enables the prediction of observables that change under rotation, such as the electron density. This is achieved by introducing rotationally equivariant atom-wise representations $\mathbf{x}_i^{(l)}\in \mathbb{R}^{2(l+1)xC}$, where the scalar ``channels'' $C$ are replaced by representations derived from spherical harmonics up to a maximum degree $L$, that now characterize atomic environments in a rotationally equivariant manner. To produce these representations, we used PhiSNet, a neural network architecture that employs equivariant atomic representations to reconstruct the electronic Hamiltonian.~\citep{unke2021se} accurately. The representation module takes as inputs nuclear charges $Z$ and positions $\mathbf{R}$ of $M$ atoms, which are used to construct equivariant spherical tensor feature representations encoding information about the chemical environment of each atom. 
The initial atomic feature representations $\mathbf{x}$ are based on element descriptors, which encode information about the nuclear charge and the ground state configuration of each element, providing an inductive bias that takes into account the quantum chemical structure of different elements (see Supplementary Note~1 for detailed building block definitions). To construct the atom-wise representations used for density prediction, the initial atomic feature representations are refined by three sequential modules, each consisting of identical building blocks with independent parameters. 
An illustration of the architecture for atomic feature representations can be found in Figure~\ref{fig:nn_architecture}.

\section*{Data Availability}
The data supporting the findings of this study are available within the paper and its Supplementary Information files. Source data for all figures are provided with this paper. Additional data are available from the corresponding authors upon reasonable request.

\section*{Code Availability}
The code used to train and evaluate the DenSNet models, generate molecular dynamics trajectories, and compute infrared spectra is available at \url{https://github.com/MihailBogojeski/equiv_dens_ml}.


\bibliographystyle{unsrtnat}
\bibliography{references}

\section*{Acknowledgements}
The authors thank the Camille and Henry Dreyfus Foundation (grant no.\ ML-22-146 to M.E.T.).
M.R.H.  was supported by the Simons Center for Computational Physical Chemistry (SCCPC) at NYU (SF Grant No. 839534).
K.R.M.\ and M.B., were partly funded by the German Ministry for Education and Research (BMBF) as BIFOLD – Berlin Institute for the Foundations of Learning and Data - under Grants 01IS14013A-E, 01GQ1115, 01GQ0850, 01IS18025A, 031L0207D, and 01IS18037A.   
K.R.M.\ was also supported by the Institute of Information \& communications Technology Planning \& Evaluation (IITP) grants funded by the Korea government (MSIT) (No.\ RS-2019-II190079, Artificial Intelligence Graduate School Program of Korea University and No.\ RS-2024-00457882, AI Research Hub Project).

\section*{Author Contributions}
M.B.\ developed the equivariant neural network architecture for density prediction and performed model training. M.R.H.\ performed molecular dynamics simulations, spectroscopic analysis, and polythiophene extrapolation experiments. L.V.-M.\ assisted with dataset generation and reference DFT calculations. K.-R.M., K.B., and M.E.T.\ conceived and supervised the project. All authors contributed to the manuscript.

\section*{Competing Interests}
The authors declare no competing interests.

\end{document}